\documentclass[letterpaper]{article} 
\usepackage{aaai23}  
\usepackage{comment}
\usepackage{times}  
\usepackage{helvet}  
\usepackage{courier}  
\usepackage[hyphens]{url}  
\usepackage{graphicx} 
\usepackage{natbib}  
\usepackage{caption} 
\usepackage[linesnumbered, ruled,vlined]{algorithm2e}

\usepackage{amsmath,amssymb,xcolor,mathtools,amsfonts,tabularx,booktabs,caption, subcaption,soul}

\newcommand{\overbar}[1]{\mkern 1.5mu\overline{\mkern-1.5mu#1\mkern-1.5mu}\mkern 1.5mu}

\newcommand{\mohammad}[1]{\textcolor{cyan}{#1}}

\newcolumntype{M}[1]{>{\centering\arraybackslash}m{#1}}

\usepackage{newfloat,float}
\usepackage{listings}
\DeclareCaptionStyle{ruled}{labelfont=normalfont,labelsep=colon,strut=off} 
\floatstyle{ruled}
\newfloat{listing}{tb}{lst}{}
\floatname{listing}{Listing}
\title{AMUSE: Adaptive Multi-Segment Encoding for Dataset Watermarking}
\author{
    ${\text{Saeed Ranjbar Alvar}}^{\rm 1}$, 
    ${\text{Mohammad Akbari}}^{\rm 1}$, 
    ${\text{David (Ming Xuan) Yue}}^{\rm 1}$, 
    ${\text{Yong Zhang}}^{\rm 1}$,      
     }
\affiliations{
    \textsuperscript{\rm 1} Huawei Technologies Canada Co. Ltd, Burnaby, Canada \\
}

\usepackage{bibentry}

\begin{document}

\maketitle

\begin{abstract}
Curating high quality datasets that play a key role in the emergence of new AI applications requires considerable time, money, and computational resources. So, effective ownership protection of datasets is becoming critical. Recently, to protect the ownership of an image dataset, imperceptible watermarking techniques are used to store ownership information (i.e., watermark) into the individual image samples. Embedding the entire watermark into all samples leads to significant redundancy in the embedded information which damages the watermarked dataset quality and extraction accuracy. In this paper, a multi-segment encoding-decoding method for dataset watermarking (called AMUSE) is proposed to adaptively map the original watermark into a set of shorter sub-messages and vice versa. Our message encoder is an adaptive method that adjusts the length of the sub-messages according to the protection requirements for the target dataset. Existing image watermarking methods are then employed to embed the sub-messages into the original images in the dataset and also to extract them from the watermarked images. Our decoder is then used to reconstruct the original message from the extracted sub-messages. The proposed encoder and decoder are plug-and-play modules that can easily be added to any watermarking method. To this end, extensive experiments are preformed with multiple watermarking solutions which show that applying AMUSE improves the overall message extraction accuracy upto 28\% for the same given dataset quality. Furthermore, the image dataset quality is enhanced by a PSNR of $\approx$2 dB on average, while improving the extraction accuracy for one of the tested image watermarking methods. 
\end{abstract}

\section{Introduction}
High quality datasets significantly contributed to the success of recent advances in the field of Artificial Intelligence (AI). As a result of rapid adoption of AI-based tools and services in different domains, the need for specialized datasets is rapidly growing~\cite{dataset_market}. The data marketplaces such as AWS Data Exchange \cite{AWS} and Snowflake Marketplace \cite{snowflake} aim to address this need by providing platforms that help owners (i.e., providers) to monetize and sell their datasets to the potential buyers (i.e, consumers). Curating such datasets requires significant time, money, computational resources, and expertise, which make them valuable intellectual assets for the stakeholders including the corporations and individuals.
Therefore, effective ownership protection of the datasets is crucial. However, the current data marketplaces only rely on user agreements to protect ownership of datasets. Since user agreements can be easily ignored and bypassed, the existing marketplaces do not offer strong ownership protection.
For instance, the buyers may claim ownership of the purchased data and sell it in other markets, or they may illegally leak the data they agreed to keep private. In such cases, existing data marketplaces neither have solid solutions to verify the data ownership nor they can trace the owner's assets. 

\begin{figure}[!t]
\includegraphics[width=0.99\columnwidth]{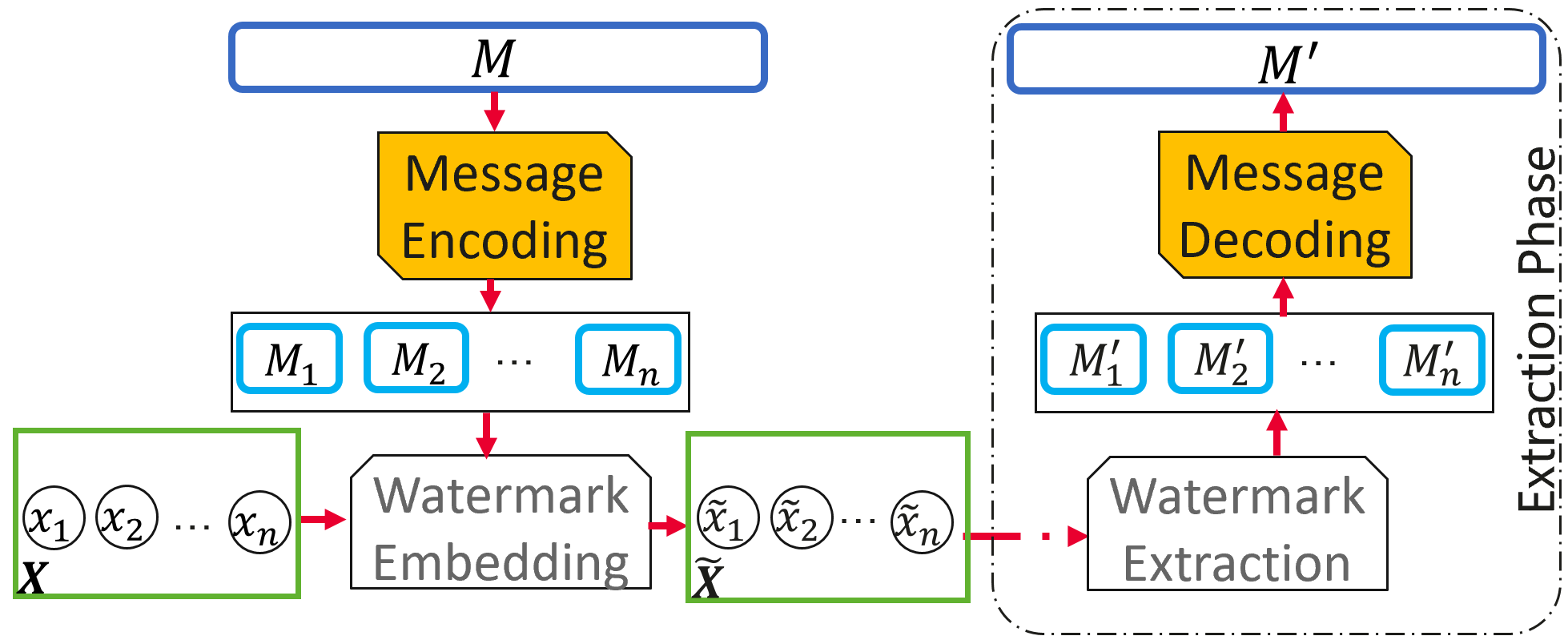}
\centering
\caption{The overview of message encoding-decoding for dataset watermarking.}
\label{fig:overview}
\end{figure}

Digital watermarking methods, which often embed a hidden signature into a digital file, are proposed to address the issues related to the ownership proof and data tracing for different data types~\cite{trailchain, kdd, drm_wm}. Among them, watermarking of image datasets has attracted significant attention in recent years, which is the focus of this paper. Image dataset watermarking methods can be divided into two groups. In the first group ~\cite{radioactive, isotopes, backdoor, clean-label, untargeted}, 
an image dataset is watermarked by modifying either the data samples or the labels that change the output of the model trained on the watermarked dataset.
Later, the model's output is analyzed to detect if it has been trained on the watermarked dataset. 

In the second group, an alternative solution to image dataset watermarking is proposed in~\cite{wm_gan}, where a learning-based imperceptible 
watermarking method is used to embed a watermark logo in every image in the dataset. Imperceptibility implies that the quality of the watermarked images should not degrade noticeably compared to the original images. Such watermarking methods have a trade-off between imperceptibility (the watermarked dataset quality), capacity (how much information is embedded), and the watermark extraction accuracy~\cite{imprect-capacity}. As a result, embedding the entire watermark into all samples leads to significant redundancy in the embedded information which decreases the imperceptibility and watermark extraction accuracy. {In other words, embedding messages with shorter length can potentially improve the extraction accuracy (especially against attacks) and the quality of the watermarked image samples.}



In this paper, a multi-segment encoding-decoding method for dataset watermarking (called AMUSE) is proposed to adaptively map the original watermark into a set of shorter sub-messages and vice versa. The overall framework is shown in Fig.~\ref{fig:overview}.
In AMUSE, the length of the sub-messages is adaptive as it is adjusted based on the dataset protection requirements. After the message encoding is completed, the sub-messages are embedded into the dataset samples using an off-the-shelf watermarking method. In the extraction phase, the same watermarking method is used to extract the sub-messages from the watermarked samples. The extracted sub-messages are then passed to our message decoder to reconstruct the original message. 
To the best of our knowledge, AMUSE is the first to propose a message encoding method that maps the original message into a set of shorter messages for dataset watermarking.
The main contributions of this paper are as follows:
\begin{itemize}
    \item{An adaptive multi-segment message  encoding  and  decoding  method is proposed which maps the watermark message into sub-messages and reconstructs  the  original  message  from  the   sub-messages, respectively.}
    \item{We demonstrate that the proposed method is plug-and-play which can effectively and efficiently be integrated with any existing image watermarking method.}
    \item{We empirically validate the fact the embedding longer watermark messages reduces the extraction accuracy.}
    \item Extensive experiments are preformed with multiple watermarking solutions which show that applying AMUSE improves the overall message extraction accuracy and watermarked dataset quality.
\end{itemize}

\section{Related Works}
\label{sec:related}

\textbf{Image Dataset Watermarking:} 
Most of the existing works for image dataset watermarking are based on modification of either the original samples and/or the labels that change the output of the model trained on the watermarked dataset. These works assume that there exists a model that is trained using the watermarked dataset, and use that model's output to trace the dataset usage which is a limiting factor for this group of works. In~\cite{backdoor, untargeted}, the labels of training data samples are changed such that the mis-labeled samples cause the model to learn a backdoor functionality. The inserted backdoor can later be used for ownership verification. In ~\cite{clean-label, radioactive, isotopes}, imperceptible perturbations (watermarks) are added to the images such that the output of any model trained on the watermarked dataset will reveal that the same dataset has been used for training. 

On the other hand, the only image dataset watermarking method in the literature that does not relay on a model trained on the watermarked dataset for watermark extraction is~\cite{wm_gan}. In this work, a Generative Adversarial Networks (GAN)-based method is proposed to embed a fixed watermark message into all the samples in the dataset. A combination of distortion and perceptual loss functions are used for training the model. In other word, the model is trained assuming that every image in the dataset needs to carry the entire message, which adds undesired redundancy to the images when owners are not concerned about the protection on a single image.

\textbf{Database Watermarking:} 
Database watermarking can also be considered as another category of related works that is different from dataset watermarking. Databases are generally used to store numerical or categorical data in tables with rows and columns, which require different watermarking techniques. 
In database watermarking methods~\cite{db1, db2, db3, db5}, the amount of allowed distortion due to watermarking is smaller compared to the images. As a result, the watermarks are not embedded into individual records. Instead, the database is split into segments, and individual bits of the watermark message are embedded into the obtained segments of the database. So, embedding a long watermark in a database which has limited number of tuples is infeasible. Unlike the conventional database watermarking approaches, ~\cite{db4} employed an image watermarking method to watermark numerical datasets. In this work, the attribute with lowest variance is normalized and treated as an image. Then, Singular Value Decomposition (SVD)-based image watermarking technique is used to embed the watermark message into the values of the normalized attribute. 


\section{Problem Definition} 
Dataset watermarking is defined as a method to embed message $\mathbf{M} = \{0,1\}^L$ into a given dataset $\mathbf{X} = \{{x_i}\}_{i=1}^n$ where $x_i$ is the $i$-th sample in the dataset and $n$ is the total number of samples in $\mathbf{X}$. The watermarked dataset is defined as $\mathbf{\tilde{X}} = \{{\tilde{x}_i\}}_{i=1}^n$ in which all or a subset of the samples may be affected and modified.

Ideally, a dataset watermarking method should achieve three objectives. First, the quality of the watermarked dataset should not drop significantly compared to the original dataset. The second objective is to be able to extract the embedded message correctly when a subset of the watermarked dataset is disclosed. Finally, the watermarking method should be able to embed a high capacity of bits. 

If watermark embedding and extraction methods are defined as $W$ and $\Psi$, respectively, a dataset watermarking method provides a solution to the following problem:
\begin{equation}
\label{eq:problem}
\begin{aligned}
\min_{W,\Psi} \quad & e_{w}  (\Psi ( W (\mathbf{X},M)),M) \\
\textrm{s.t.} \quad & e_{d}(\mathbf{\tilde{X}},\mathbf{{X}})< \epsilon_d,  \quad \tau \leq \hat{\tau}, \quad I\geq \hat{I},
\end{aligned}
\end{equation}
where $W (\mathbf{X},M)$ embeds watermark message $M$ into dataset $\mathbf{X}$,  
and $\Psi(.)$ extracts the embedded message from the 
watermarked dataset. $e_w$ is the error function used to measure the error in the extracted message compared to the original message $M$. $e_d$ is another error function to measure the distortion (i.e., used for evaluating the watermarked dataset quality) caused due to watermarking. 
$\epsilon_d$ is a predefined threshold on dataset distortion which defines how much degradation in quality is allowed in the watermarked dataset. ${\tau}$ denotes the ratio of the original dataset, wherein the disclosure of any subset of dataset exceeding the size of ${\tau}$ ensures a guaranteed watermark extraction. $\hat{\tau}$ is a predefined threshold that defines what subset of watermarked dataset is required to guarantee correct message extraction. Finally, $I$ is the capacity of embedded information in the watermarked dataset, and $\hat{I}$ is a predefined threshold which sets the minimum message length in bits.

\section{Proposed Method}
\label{sec:proposed}
In this section, 
the proposed message encoding-decoding approach (called AMUSE) for dataset watermarking is discussed to solve the problem defined in the previous section. We first introduce the adaptive multi-segment encoder that is used to map the original message into a set of sub-messages. Then, a dataset watermarking method based on the introduced encoder is presented. Finally, the details of the message decoding procedure which is employed for reconstructing the original message are provided (Fig.~\ref{fig:overview}).

In our dataset watermarking framework, rather than embedding the entire message into all the samples, we distribute the message over a group of samples using the proposed message encoder. The proposed dataset watermarking method relies on the fact that a dataset is often valuable when more than one sample is available, and if individual samples are leaked, they would not be as useful as the larger sets of data. Therefore, sample level protection can be relaxed. 

\subsection{Adaptive Multi-Segment Encoder}
\label{subsec:prop_encoder}
The message encoder, denoted by $E$, maps the original message $M$ into a set of equal-length sub-messages 
$\overbar{M}=\{M_i\}_{i=1}^{n}$ where $M_i$ is the sub-message to be embedded into the data sample $x_i$. The length of each sub-message $|M_i|=l$ must be less than or equal to the original message's length defined as $L = |M|$ (i.e., $l\leq L$). The encoder splits the original message into $N$ chunks, and creates sub-messages by excluding $K$ chunks. We will later discuss $N$ and $K$ selection and how to use them to define the sub-message length.

\begin{figure}[t]
\includegraphics[width=\columnwidth]{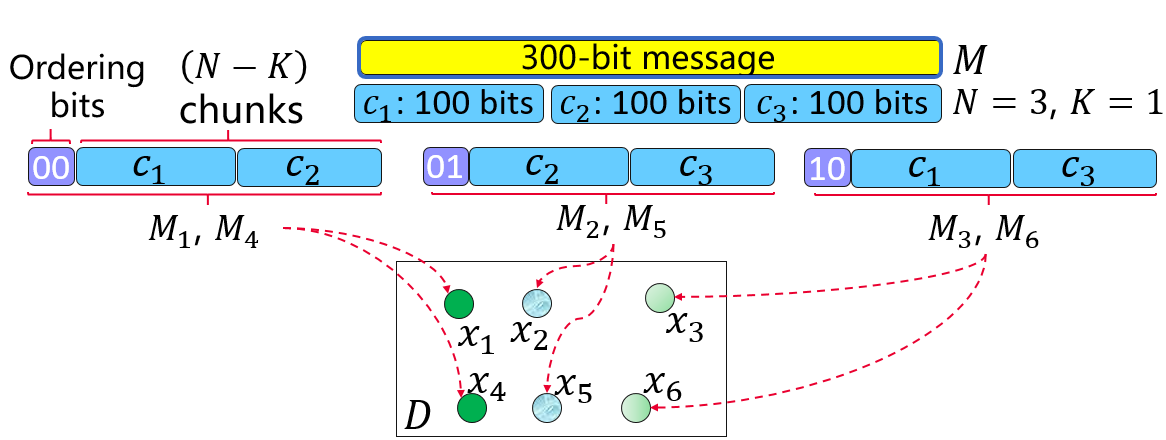}
\centering
\caption{An example for encoding a 300-bit message with $N=3$, $K=1$, and $n=6$. The length of the obtained sub-messages is 202 bits.}
\label{fig:enc_example}
\end{figure}

If the original message length is not divisible by the number of chunks, it is zero padded by $p$ bits. Then, the padded message is split into $N$ chunks $c =  \{c_1, \dots, c_N\}$. Next, $C=\binom{N}{N-K}$ combinations of chunks, denoted by $S= \{s_1, \dots,  s_C \}$, are obtained. The combinations are then updated by concatenating the ordering bits that are used to indicate the combination index. The combination index is later used for correct extraction of the original message. Finally, the set of sub-messages $\overbar{M}$ is obtained by assigning the {combined chunks and their ordering bits} to the samples of the dataset. 
The step-by-step encoding procedure is shown in Algorithm~\ref{alg:message-encode}. {It is worth noting that in Step 7 of Algorithm~\ref{alg:message-encode}, we generate a message for each sample in the dataset by reusing the chunks obtained in Step 6. This step can be performed dynamically during the watermark embedding process, so we do not need to store the messages for all the samples in the dataset. This improves the run time and the memory requirement of the encoder for large-scale datasets.} An example for encoding a 300-bit message into six 202-bit messages is also shown in Fig.~\ref{fig:enc_example}.

\SetKwInput{KwData}{Input}
\SetKwInput{KwResult}{Return}
\begin{algorithm}[tb!]
\small
    \caption{Message Encoding Algorithm}\label{alg:message-encode}
    \KwData{$M, N, K, n$, $\oplus$ (the concatenation operator), $B$ (the function to convert an integer index to binary representation)} 
    \If {\(L \bmod N \neq 0\)}
    {
        Pad \(M\) with \(p = N - (L \bmod N)\) zeros\
        }
    Split \(M\) into \(N\) chunks, \(c = \{c_1, \dots, c_N\}\) \\
    Obtain all the $N-K$ combination of chunks as $S=\{s_1,\dots, s_C\}$, where $C= \binom{N}{N - K}$\\ 
    Let \(\phi:  S \rightarrow \{1,\dots,C\}\) be bijective mapping to assign an index to the combinations in $S$ \\
    \(S' = \{B(\phi(i)) \oplus S[i] : 1 \leq i \leq C\}\)\\
    \(\overbar{M} = \{S'[C-(j \bmod C)] : 1 \leq j \leq n\}\)\\
    \KwResult{\(\overbar{M}\), $p$}
\end{algorithm}



\subsubsection{$N$ and $K$ selection:}
$N$ and $K$ values define the length of the sub-messages and the number of required samples to reconstruct the original watermark message from the {sub-messages}. 
The shortest {sub-message} length that can be embedded per sample is one plus the number of ordering bits. 
However, for this case, we need at least $L$ samples to reconstruct the original message. When $L$ gets larger, more samples are required to reconstruct the original message. Requiring large number of watermarked samples to extract the original message makes dataset watermarking solution prone to subset attacks. In a subset attack, adversary leaks a subset of the dataset such that the watermark cannot be extracted from the leaked samples.  
Indeed, there is a trade-off between the embedded message length and the robustness against subset attacks. Hence, $N$ and $K$ are obtained by solving the following problem which considers both the {sub-}message length and the robustness against such attacks: 
\begin{equation}
\label{eq:len_op}
\begin{aligned}
\min_{N,K} \quad & l  (N,K), \quad
\textrm{s.t.}  \quad  \tau \leq \hat{\tau},
\end{aligned}
\end{equation}
where $l$ is the length of the {sub}-messages, i.e., $l=|M_i|$. ${\tau}$ denotes the ratio of the original dataset, where the disclosure of any subset from the dataset larger than ${\tau}$ guarantees a successful message extraction, and $\hat{\tau}$ is a predefined threshold which is set by the dataset owner. For the encoding procedure presented earlier (Algorithm~\ref{alg:message-encode}), the sub-messages include $N-K$ chunks out of $N$ chunks. The worst case scenario for a subset of the dataset for which original message cannot be extracted is when all chunks of the original message are not included in the extracted sub-messages. In other words, at least 1 chunk is missing in the sub-messages that are extracted from the given watermarked subset. Hence, ${\tau}$ for our proposed encoder is $\tau = \frac{\binom{N-1}{N-K}}{\binom{N}{N-K}}$ that gives $\tau = \frac{K}{N}$.

For any subset of the {watermarked} dataset with a ratio larger than $\tau$, the {embedded} sub-messages are guaranteed to have all the chunks {needed for reconstructing} the original message. 
This holds as long as the extracted sub-messages from the subset are error-free. For the extracted messages with error, $\tau$ is expected to be larger.

If $\tau = \frac{K}{N}$ is replaced in Eq. \eqref{eq:len_op}, the optimization problem in \eqref{eq:len_op} can be solved by a brute-force search over the range of $N$ and $K$ and computing the length by applying the encoder $E$ to the {original} messages. The set of test values for $N$ and $K$ can be reduced by choosing the sets that are suitable for practical applications. For instance, we assume that the minimum leaked ratio should be 1\% for a dataset. The details of the algorithm to solve Eq. \eqref{eq:len_op} is given in Algorithm~\ref{alg:optimal-select}.   

\begin{algorithm}[tb!]
\small
    \caption{Optimal $N$ and $K$ selection algorithm}\label{alg:optimal-select}
    \KwData{$\hat{\tau}, M, n$, \(E\) (the message encoding algorithm)}
        \(l^* = L\), \(N^* = 1\), \(K^* = 0\)\\
        \For{\(N \in 2 \dots 100\)}{
            \For{\(K \in 1 \dots N-1\)}{
                \(\tau = \frac{K}{N}\)\\
                \If {\(\tau > \hat{\tau}\)}{
                    Continue
                }
                \If {\(\binom{N}{N-K} > n\)}{
                    Continue
                }
                \(\overbar{M} = \{M_i\}^{n}_{i = 1} = E(M, N, K,n)\)\\
                \If{\(|M_i| > l^*\)}{
                    Continue
                }
                \(l^* = |M_i|\), \(N^* = N\), \(K^* = K\)\\
            }
        }
    \KwResult{\(N^*, K^*\)}
\end{algorithm}

AMUSE is an adaptive method that adjusts the protection level according to the needs of the users. Specifically, the embedded message length changes according to the given $\hat{\tau}$. If a small subsets of the dataset are as important as the entire dataset, then more message chunks are embedded into the sub-messages, whereas, the embedded chunks are reduced for larger thresholds such as $\hat{\tau}>50\%$.

\subsection{Dataset Watermarking using AMUSE}
\label{subsec:propo_wm}
AMUSE can be integrated with the existing watermarking solutions. Specifically, the original message is encoded with AMUSE and the samples of the dataset are watermarked with the sub-messages obtained from AMUSE. 
In this paper, we utilize off-the-shelf watermarking methods to embed the sub-message $M_i$ into the $i$-th sample of the dataset, i.e., $x_i$. 
The watermarking process is repeated for all the samples (along with their corresponding sub-messages) in the dataset to obtain the final watermarked dataset. 
Since the focus of this paper is on image datasets, conventional or learning-based image watermarking solutions {such as~\cite{dctdwt},~\cite{hidden}, and ~\cite{ssl}} can be used to watermark the images in the dataset. 

Note that our proposed message encoding is independent from the off-the-shelf watermarking methods in obtaining the sub-messages. Therefore, it is a plug-and-play module that can be integrated with any given watermarking method. 

The extraction module in the off-the-shelf watermarking method is also used to extract the embedded sub-messages from the available samples of the watermarked dataset. 
The available watermarked dataset, denoted by $\mathbf{\tilde{X}_s}$, is indeed a subset from the watermarked dataset (i.e., $\mathbf{\tilde{X}_s} \subset \mathbf{\tilde{X}}$).  The set of extracted sub-messages is denoted as $\overbar{M}^{'}=\{M_i^{'}\}_{i=1}^{m}$, where each extracted sub-message $M_i^{'}$ includes some chunks of the original message rather than the entire message. $m$ is the number of samples in $\mathbf{\tilde{X}_s}$ (i.e., $m \leq n$). 

\subsection{AMUSE Decoder}
Given the extracted sub-messages $\overbar{M}^{'}$, AMUSE decoder $D$ is employed to reconstruct the original message. $D$ basically reverses the encoding procedure performed by $E$. First, the ordering bits are obtained from all extracted sub-messages $\{M_i^{'}\}_{i=1}^{m}$. We assume that $N$, $K$, and the number of padding bits $p$ are known for a given watermarked dataset.  So, the number of ordering bits is known for the extraction service. In addition, we assume that the corresponding combination of chunks for the given ordering index is known. Therefore, the extracted ordering bits can be used to group all the same chunks that are in the extracted sub-messages. Next, majority voting is applied to the {grouped} chunks to obtain the final chunks. The final chunks are merged together and the $p$ padding bits are discarded to reconstruct the final reconstructed message $M'$. If there is no error in the extraction, all the bits in the extracted message are equal to the bits in the original message. 

\section{Experimental Results}
\label{sec:exp}
{
In this section, the performance of the proposed AMUSE method for image dataset watermarking is numerically evaluated and compared with the baselines. In the following sub-sections, the experimental settings, metrics, impact of message length ($L$), effectiveness and generality of AMUSE as a plug-and-play module, and its robustness against different attacks including subset attacks are presented.
}

\subsection{Experimental Settings} 
In order to analyze the performance and generality of AMUSE, two learning-based image watermarking approaches including HiDDeN \cite{hidden} and SSL \cite{ssl}, and a frequency-based method named DCT-DWT \cite{dctdwt} are used for our experiments. The mentioned off-the-shelf methods are indeed employed to perform "Watermark Embedding" and "Watermark Extraction" over individual images in the dataset (Fig.~\ref{fig:overview}). 

For the SSL method, the original implementation with the default arguments in \cite{ssl} was utilized. For HiDDeN, we used the Pytorch implementation\footnote{https://github.com/ando-khachatryan/HiDDeN} and trained their model with different message lengths for 400 epochs (details are given in the {Appendix}). For DCT-DWT, the implementation in the invisible-watermark
repo\footnote{https://github.com/ShieldMnt/invisible-watermark} was used. For learning-based methods, the experiments are done on TITAN XP GPUs.
For SSL and DCT-DWT, 10 $L$-bit original messages are randomly generated for each $L\in \{100, 200, 300\}$. 
The training procedure for the same message lengths did not converge for the HiDDeN. Therefore, we used shorter message lengths with $L\in \{30,16,9\}$. 


For all our experiments, a dataset of 100 random images from ImageNet validation set is chosen. For training HiDDeN, 2.6K images from ImageNet train set are used. As in \cite{hidden}, center cropping of size $128\times128$ is applied for HiDDeN in both training and evaluation.

{The only work in the literature for image dataset watermarking is \cite{wm_gan}. However, we did not provide comparison results with this method as it is not in the scope of our work because: 1) it only works with a single image logo as watermark (directly provided as input to their CNN model), and 2) for each different logo, they needed to train separate models (unlike our method that works with any watermark with no extra training). }

{Although there is no related work to be compared with our method, we design some baselines based on the HiDDeN, SSL, and DCT-DWT methods. For the baselines in all the experiments in this paper, we embed the entire original message to all the images in the dataset. To ensure a fair and consistent comparison with the proposed method, 
majority voting is applied to the extracted messages from the watermarked dataset samples for all the baselines to obtain the final reconstructed message. It should be noted that \textit{the same watermark message} is embedded into and extracted from the dataset for the baselines and the proposed method.}



\subsection{Metrics}
Two metrics are used for evaluating the accuracy of the extracted messages. The first metric is Bit Accuracy (BA)~\cite{hidden} that is defined as: 
\begin{equation}
\label{eq:ba}
     BA = \frac{1}{L} \sum_{j=1}^{L} \neg(M'^{j}\otimes M^j),
\end{equation}
where $\otimes$ is XOR operator, $\neg$ is logical negation, $M'^{j}$ is the $j$-th bit in the reconstructed message $M'$, and $M^{j}$ is the $j$-th bit in the original message $M$. The second metric is Word Accuracy (WA), which evaluates if the entire extracted message is error free~\cite{word_accuracy}, is defined as: 
\begin{equation}
\label{eq:wa}
  WA =
    \begin{cases}
      1 & if \quad $BA==1$  \\
      0 & \text{otherwise}.
    \end{cases}       
 \end{equation}


Peak Signal-to-Noise Ratio (PSNR) and average PSNR are used to measure the qualities of the watermarked images and the watermarked dataset, respectively.

\subsection{Does $L$ matter in dataset watermarking?}
We study the effect of the watermark message length $L$ on the extraction accuracy with the HiDDeN- and SSL-based dataset watermarking for both no-attack and with-adversarial-attack (referred as with-attack in the remainder of the paper) scenarios. 
Note that in the following experiments, AMUSE is not applied and the original message is embedded into all the samples of the dataset.

\begin{figure}[!tb]
  \begin{subfigure}[t]{0.47\columnwidth}
    \includegraphics[width=\columnwidth]{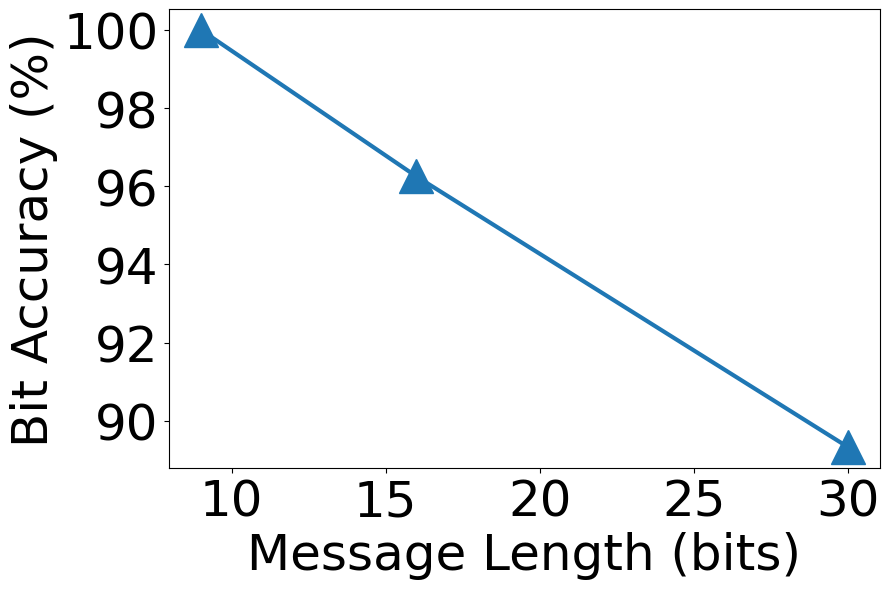}
    \caption{HiDDeN, No attack}
    \label{fig:hidden_l_exp1}
  \end{subfigure} 
  \begin{subfigure}[t]{0.47\columnwidth} 
    \includegraphics[width=\columnwidth]{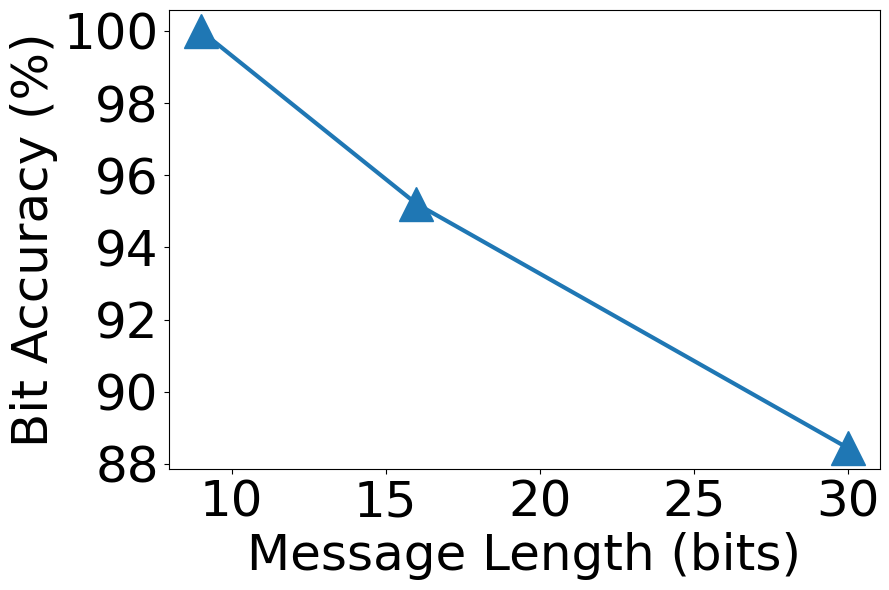}
    \caption{HiDDeN, with attacks}
    \label{fig:hidden_l_exp2}
  \end{subfigure}
  \begin{subfigure}[t]{0.47\columnwidth}
    \includegraphics[width=\columnwidth]{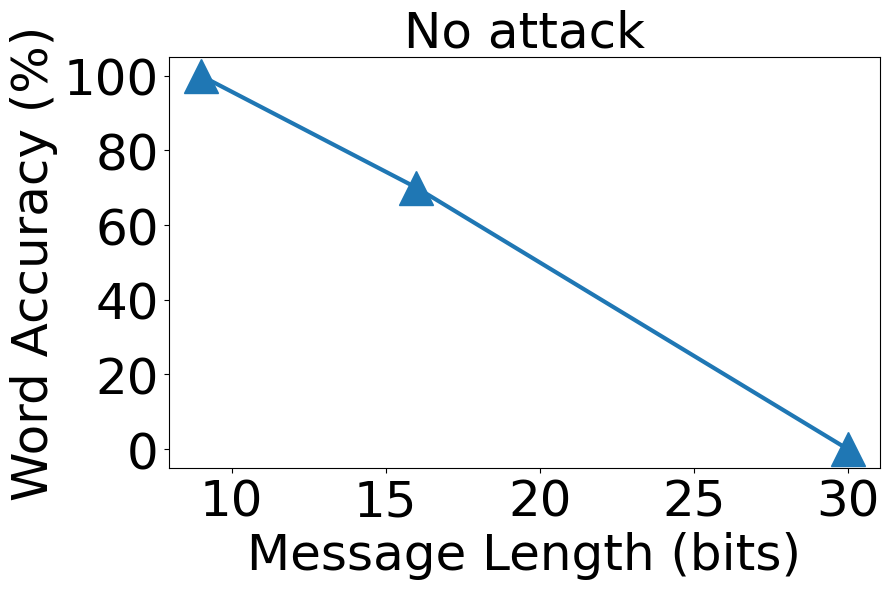}
    \caption{HiDDeN, No attack}
    \label{fig:hidden_l_exp3}
  \end{subfigure} 
  \begin{subfigure}[t]{0.47\columnwidth} 
    \includegraphics[width=\columnwidth]{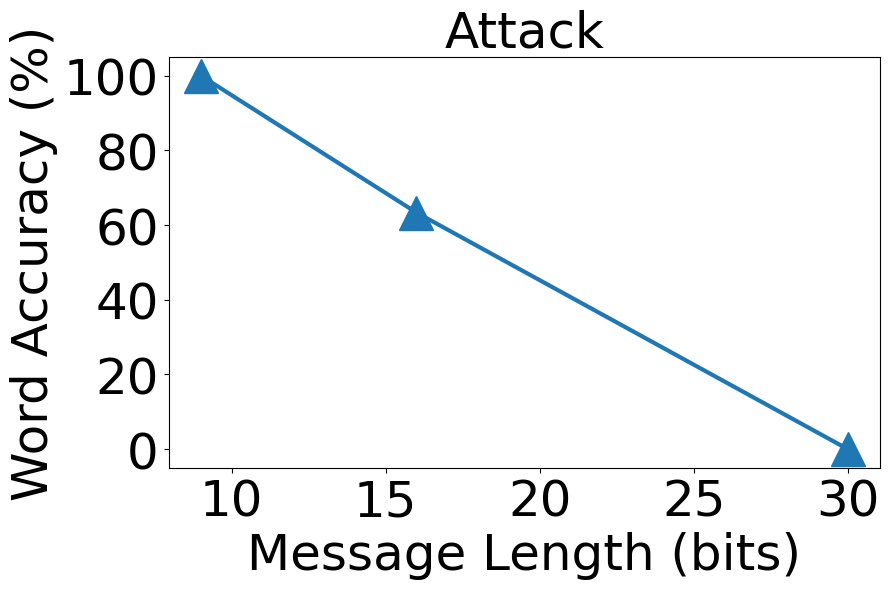}
    \caption{HiDDeN, with Attacks}
    \label{fig:hidden_l_exp4}
  \end{subfigure}  
\caption{Extraction bit (top) and word (bottom) accuracy vs. message length for HiDDeN-based dataset watermarking.}
\label{fig:msg_l_hidden_ba}
\end{figure}

\textbf{HiDDeN:} For this method, we trained 3 different models for 3 message lengths $L \in \{30,16,9\}$. Following the original work in \cite{hidden}, data augmentations corresponding to a set of pre-defined attacks are applied during the training. We then tested the model with the attacks that were considered for data augmentation during the training (details in the {Appendix}). The experiments are repeated with 10 different messages and the average bit and word accuracy vs. message length are shown in Fig.~\ref{fig:msg_l_hidden_ba}. As shown in the figure, the bit extraction accuracy for both no-attack (Fig.~\ref{fig:msg_l_hidden_ba}-a) and with-attack (Fig.~\ref{fig:msg_l_hidden_ba}-b) cases is higher when the embedded message is shorter. 
As shown in Fig.~\ref{fig:msg_l_hidden_ba}-c and Fig.~\ref{fig:msg_l_hidden_ba}-d, similar to the bit accuracy, word accuracy is also higher when the embedded message is shorter. Zero word accuracy for 30-bit messages means that there is at least one bit error in all the extracted messages. 

\begin{figure}[!tb]
\centering
\includegraphics[width=0.477\columnwidth]{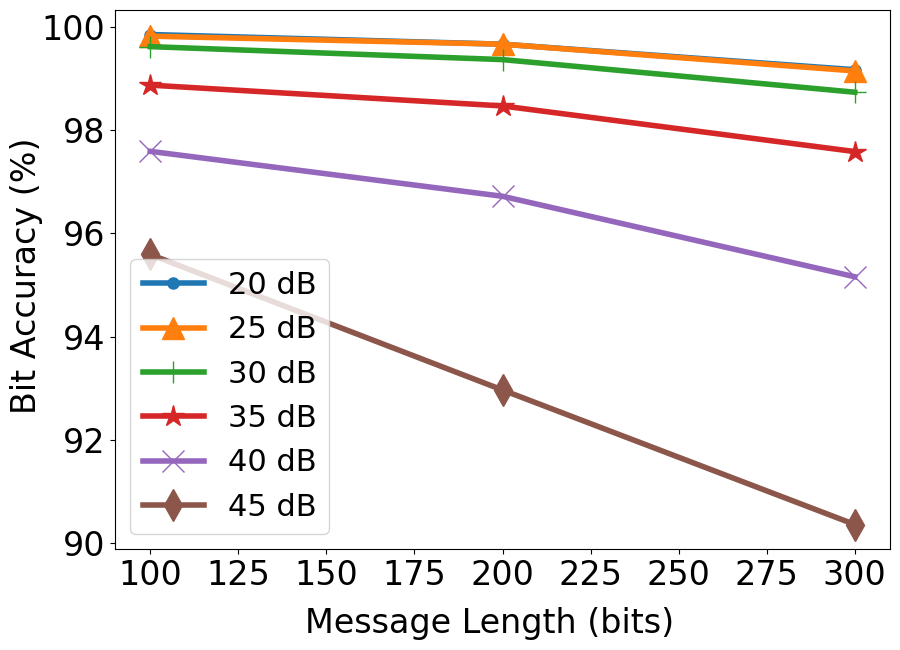}
\includegraphics[width=0.513\columnwidth]{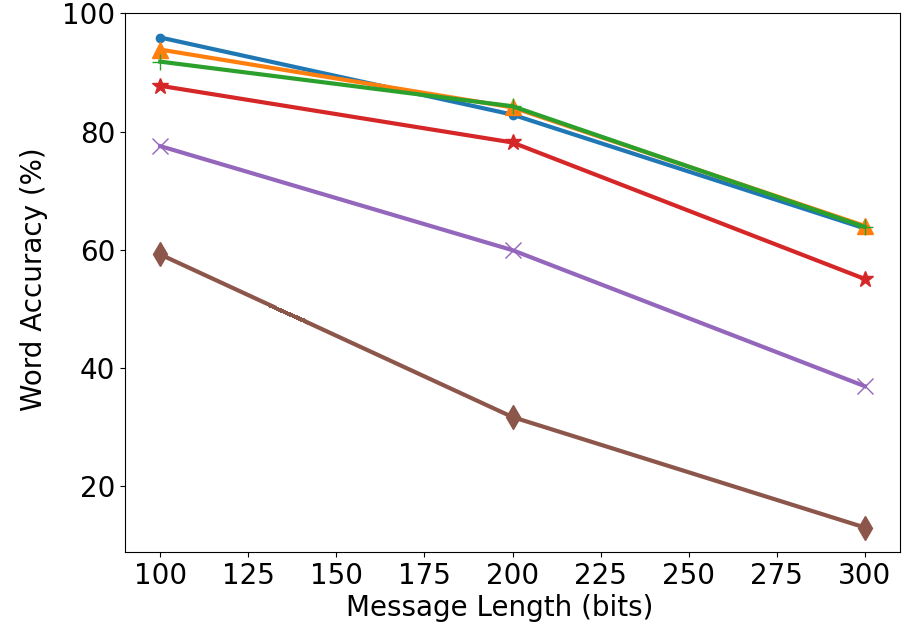}
\caption{Average bit and word accuracy (with attacks) vs. message length for the SSL-based dataset watermarking.}
\label{fig:msg_l_ssl_ba}
\end{figure}

\textbf{SSL:} We also evaluate the effect of the message length on the extraction accuracy with the SSL method. As mentioned earlier, unlike HiDDeN, SSL 
can handle longer messages. Therefore, we tested message lengths $L\in \{100,200,300\}$ and generated 10 random messages for each length. SSL is also robust to wider range of adversarial attacks. Hence, it is evaluated with 45 different geometric and non-geometric attacks such as blur, crop, resize, rotation, color changes, and JPEG compression (more details in the {Appendix}).

In contrast to HiDDeN, SSL allows the noise intensity to be adjusted for watermarked images according to a target PSNR. 
Therefore, the average bit and word accuracy vs. the message length are obtained for different PSNR values. 
SSL achieves 100\% bit accuracy when no adversarial attack is applied, but as expected, the bit accuracy drops after performing adversarial attacks. The average bit and word accuracy over the attacks are shown in Fig.~\ref{fig:msg_l_ssl_ba}. Higher PSNR means that the smaller perturbation is allowed to be added in the watermarked data. 

As shown in Fig.~\ref{fig:msg_l_ssl_ba}, the bit accuracy drops when the message length becomes larger. The results in the figure also show that the drop in the bit accuracy is larger for higher target PSNR values. As for HiDDeN, the average word accuracy with SSL has a similar trend, where it drops when the embedded message length increases.

\subsection{Does applying AMUSE help?}
Next, we evaluate the effect of applying AMUSE on the extraction accuracy and the quality of the watermarked datasets. 
In this experiment, HiDDeN and SSL are used as off-the-shelf watermarking methods. 

\textbf{HiDDeN + AMUSE:} {For the HiDDeN-based scenario, following~\cite{hidden}, the embedded message is set to $L=30$. 
For AMUSE, $N$ and $K$ are chosen according to the subset attack thresholds $\hat{\tau}=60\%$ and $\hat{\tau}=80\%$ (Algorithm~\ref{alg:optimal-select}), which result in $(N,K)=(5,3)$ and $(N,K)=(5,4)$, respectively. Then, the proposed message encoding (Algorithm \ref{alg:message-encode}) is applied to obtain the sub-messages. The length of the obtained sub-messages are 16-bit and 9-bit for $\hat{\tau}=60\%$, and $\hat{\tau}=80\%$, respectively.}

We use the HiDDeN's trained models to embed the sub-messages into the original images of the dataset. 
The embedded sub-messages can then be extracted from the watermarked dataset for both no-attack and with-attack scenarios. 

The experiments are repeated with 10 different messages of length $L=30$. The average bit and word accuracies with and without adversarial attacks are summarized in Table \ref{tbl:amuse_hidden}. {Note that the WA of 0 for the baseline means that there is at least one bit error in all the reconstructed messages.} The detailed list of attacks is given in the {Appendix}. 

\begin{table}[!tb]
    \small
    \centering
    \begin{tabular}{M{2.2cm} M{0.84cm} M{0.70cm} M{0.80cm} M{0.70cm} M{0.75cm}}
    \toprule
        ~ & \textbf{BA} & \textbf{WA} & \textbf{PSNR} & \textbf{BA*} & \textbf{WA*} \\ \hline \hline
        HiDDeN \\(30 bits/sample) & 89.33 & 0 & 31.78 & 88.55 & 0 \\ \hline
        HiDDeN+AMUSE \\(16 bits/sample) & 99.67 & 90.00 & 32.95 & 99.22 & 85.0 \\ \hline
        HiDDeN+AMUSE \\(9 bits/sample) & \textbf{100} & \textbf{100} & \textbf{33.71} & \textbf{100} & \textbf{100} \\ \bottomrule
    \end{tabular}
    \caption{Message extraction accuracy (\%) for HiDDeN+AMUSE compared to the baseline. \textbf{BA}: Bit accuracy, \textbf{WA}: Word accuracy, \textbf{*}: with-attack.}
    \label{tbl:amuse_hidden}
\end{table}

Since HiDDeN does not work toward a given PSNR, the distortion level in the watermarked images varies for the models trained with different message lengths. 
The average PSNR for the watermarked dataset (without attacks) are given in Table~\ref{tbl:amuse_hidden}. As results indicate, when AMUSE is applied with the shortest embedded message, the average PSNR is increased by $\approx$2 dB. In other words, the watermarked dataset quality is less distorted when AMUSE is used, which is very helpful for applications such as medical image processing with the importance of less distortion. In overall, AMUSE improves both the extraction accuracy and the quality of the watermarked dataset.

Some visual examples and another set of experiments over the Oxford Flower Dataset~\cite{flower} using HiDDeN+AMUSE are given in the {Appendix}. Moreover, the effect of watermarking a training dataset using HiDDeN+AMUSE on a ML task performance is also studied and reported in the {Appendix}. 

\begin{figure}[!tb]
  \begin{subfigure}[t]{0.495\columnwidth}
    \includegraphics[width=\columnwidth]{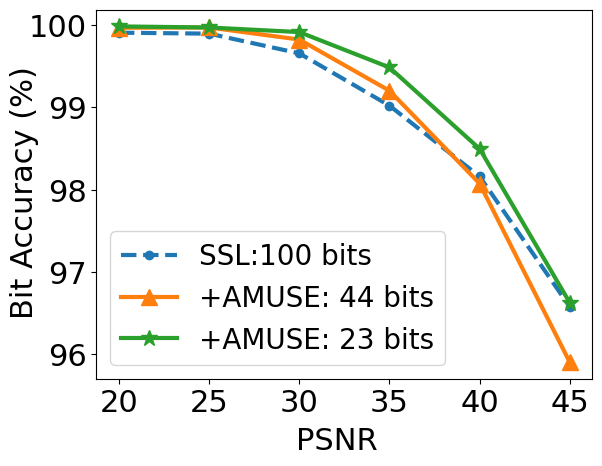}
  \end{subfigure} 
  \begin{subfigure}[t]{0.495\columnwidth} 
    \includegraphics[width=\columnwidth]{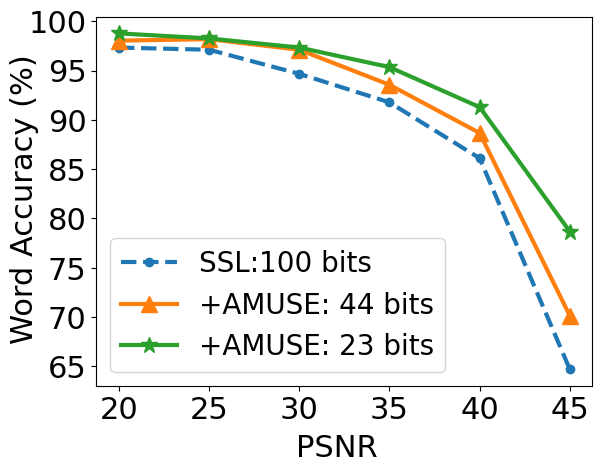}
  \end{subfigure}
  \begin{subfigure}[t]{0.495\columnwidth}
    \includegraphics[width=\columnwidth]{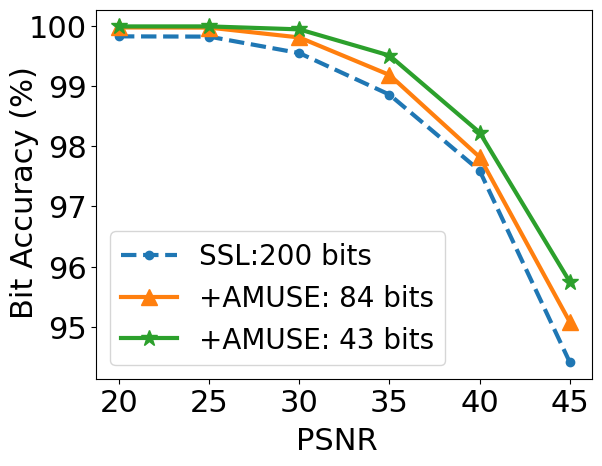}
  \end{subfigure} 
  \begin{subfigure}[t]{0.495\columnwidth} 
    \includegraphics[width=\columnwidth]{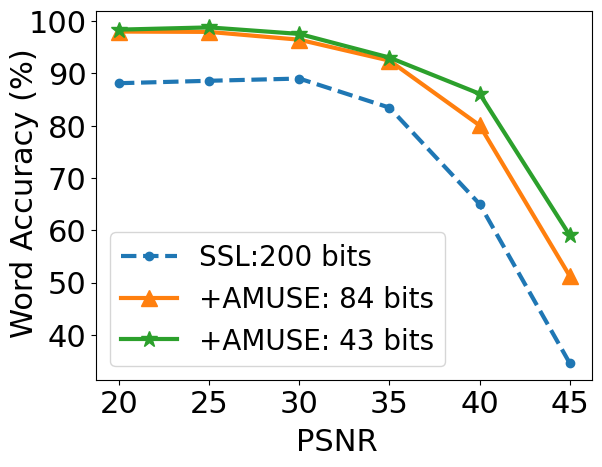}
  \end{subfigure}
  \begin{subfigure}[t]{0.495\columnwidth}
    \includegraphics[width=\columnwidth]{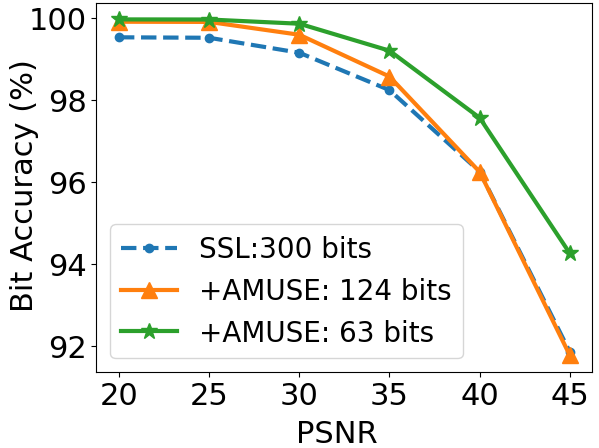}
  \end{subfigure} 
  \begin{subfigure}[t]{0.495\columnwidth} 
    \includegraphics[width=\columnwidth]{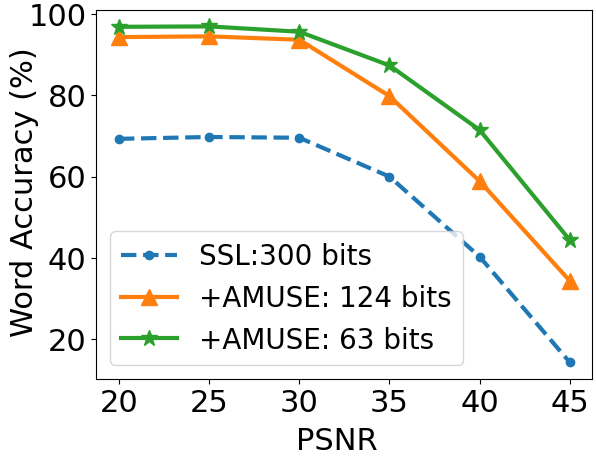}
  \end{subfigure}     
\caption{The average bit (left) and word (right) accuracy for SSL+AMUSE compared to the baseline for given PSNR values with $L=100$ (top), $L=200$ (middle), and $L=300$ (bottom).}
\label{fig:amus_ssl_ba_300}
\end{figure}

\textbf{SSL + AMUSE:} The above experiment is repeated by replacing HiDDeN with SSL in our dataset watermarking framework. As mentioned before, SSL method can handle longer messages and it is robust to more attacks compared to HiDDeN. Therefore, we tested SSL with longer messages $L=\{100,200,300\}$ and more attacks (i.e., 45 attacks). SSL can also adjust the watermarking noise level according to a given target PSNR, which results in a watermarked dataset with a predefined quality. However, the extraction accuracy after applying attacks is different for the tested PSNR values. Fig.~\ref{fig:amus_ssl_ba_300} presents the average bit and word accuracies with attacks obtained for 10 random messages of size $L=\{100,200,300\}$ bits with 6 PSNR values $\{20,25,30,35,40,45\}$. 
In this experiment, $\hat{\tau}=60\%$ and $\hat{\tau}=80\%$ are tested, which respectively correspond to {23- and 44-bit sub-messages for $L=100$, 43- and 84-bit sub-messages for $L=200$, and} 63- and 124-bit sub-messages {for $L=300$}. 
As seen in Fig.~\ref{fig:amus_ssl_ba_300}, the proposed method outperforms the baseline in terms of average word and bit accuracy for all tested $L$ and the PSNR values. 

\subsection{Is AMUSE plug-and-play?}
Since the message encoding and decoding stages are decoupled from the rest of the components in our dataset watermarking framework (Fig~\ref{fig:overview}), AMUSE can work with any existing watermarking method. Earlier, we showed the results of integrating AMUSE with 2 learning-based solutions. In this section, we study the effect of AMUSE on DCT-DWT~\cite{dctdwt} that is a frequency-based watermarking method. While DCT-DWT achieves high extraction accuracy with no attacks, its performance severely drops to random guess accuracy for any type of attack. 
Therefore, in this experiment, we focus on the results without attack. 

In DCT-DWT, a parameter called "scale" is used to control the watermarked image quality. 
With a smaller scale, less watermarking noise is added and the watermarked data quality increases (i.e., higher PSNR). In this experiment, we obtained the results using scales of 36, 30, and 20 for message lengths $L=\{100,200,300\}$. We tested DCT-DWT+AMUSE with $\hat{\tau}=60\%$, and $\hat{\tau}=80\%$.
The average results over 10 random messages are shown in Table~\ref{tbl:dwt_300}. 

As shown in the table, For $L=300$, the bit accuracy of the baseline drops over 20\% and 30\% for the scales of 30 and 20, respectively. However, when AMUSE is applied, less than 1\% and 4\% performance loss are observed for the same scales. This shows the generality of AMUSE in improving the extraction accuracy and the quality of watermarked image datasets even for conventional methods such as DCT-DWT. 
For $L=\{100,200\}$, the results indicate that bit accuracy drops up-to over 30\% for the baseline when the quality of the watermarked dataset increases. However, when AMUSE is applied, less than 1\% drop is observed which indicates that integrating AMUSE with DCT-DWT can also improve the performance of dataset watermarking.

\begin{table}[!tb]
\small
\setlength\tabcolsep{1pt}
\begin{tabular}{c c c c c}
\toprule
\textbf{}                                                             & \textbf{$L$} & \textbf{\begin{tabular}[c]{@{}c@{}}scale = 36, \\ PSNR = 36.2\end{tabular}} & \textbf{\begin{tabular}[c]{@{}c@{}}scale = 30,\\ PSNR= 37.0\end{tabular}} & \textbf{\begin{tabular}[c]{@{}c@{}}scale= 20\\ PSNR=37.7\end{tabular}} \\ \hline \hline
\begin{tabular}[c]{@{}c@{}}DCT-DWT\\ (100 bits/sample)\end{tabular}  & 100 & 100.0 & 99.10 & 83.93\\  \hline
\begin{tabular}[c]{@{}c@{}}DCT-DWT+AMUSE \\ (44 bits/sample)\end{tabular} & 100 & 100.0  & \textbf{100.0} & \textbf{100.0} \\ \hline
\begin{tabular}[c]{@{}c@{}}DCT-DWT+AMUSE \\ (23 bits/sample)\end{tabular}  & 100 & 100.0   & \textbf{100.0}   & 99.20 \\
\hline \hline
\begin{tabular}[c]{@{}c@{}}DCT-DWT\\ (200 bits/sample)\end{tabular}  & 200 & 100.0 & 86.47 & 67.20\\  \hline
\begin{tabular}[c]{@{}c@{}}DCT-DWT+AMUSE \\ (84 bits/sample)\end{tabular} & 200 & 100.0  & \textbf{100.0} & \textbf{100.0} \\ \hline
\begin{tabular}[c]{@{}c@{}}DCT-DWT+AMUSE \\ (43 bits/sample)\end{tabular} & 200 & 100.0 & \textbf{100.0}   & 99.97 \\ 
\hline \hline
\begin{tabular}[c]{@{}c@{}}DCT-DWT\\ (300 bits/sample)\end{tabular}  & 300 & 100.0                                                                       & 79.01                                                                     & 69.50                                                                  \\  \hline
\begin{tabular}[c]{@{}c@{}}DCT-DWT+AMUSE \\ (124 bits/sample)\end{tabular} & 300 &  100.0                                                                       & 99.60                                                                     & \textbf{97.55}                                                                  \\ \hline
\begin{tabular}[c]{@{}c@{}}DCT-DWT+AMUSE \\ (63 bits/sample)\end{tabular}  & 300 &  100.0                                                                       & \textbf{99.93}                                                                     & 96.61                                                                  \\ 
\bottomrule
\end{tabular}
\caption{The bit accuracy (\%) for DCT-DWT based dataset watermarking method with $L=\{100, 200, 300\}$.}
\label{tbl:dwt_300}
\end{table}

\begin{figure}[!tb]
\centering
\includegraphics[width=\columnwidth]{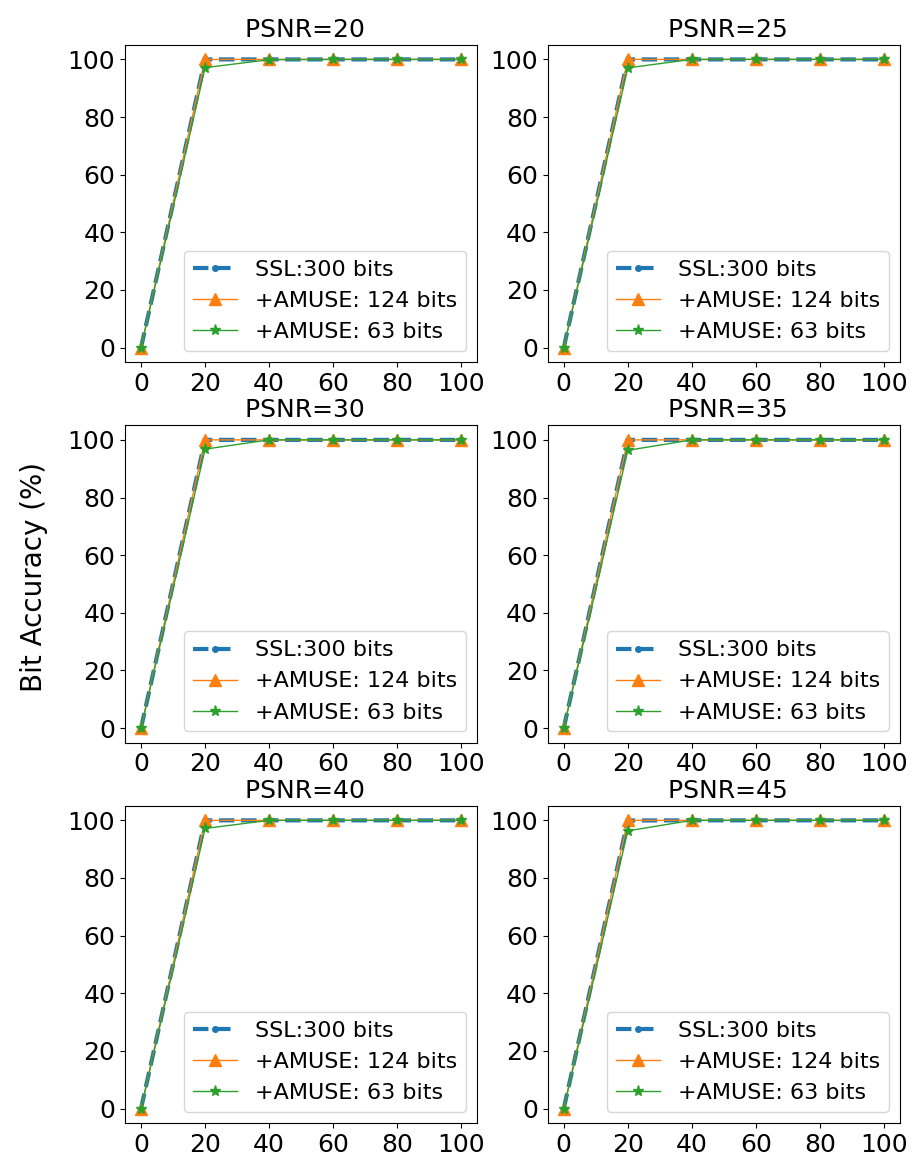}
\includegraphics[width=0.45\columnwidth]{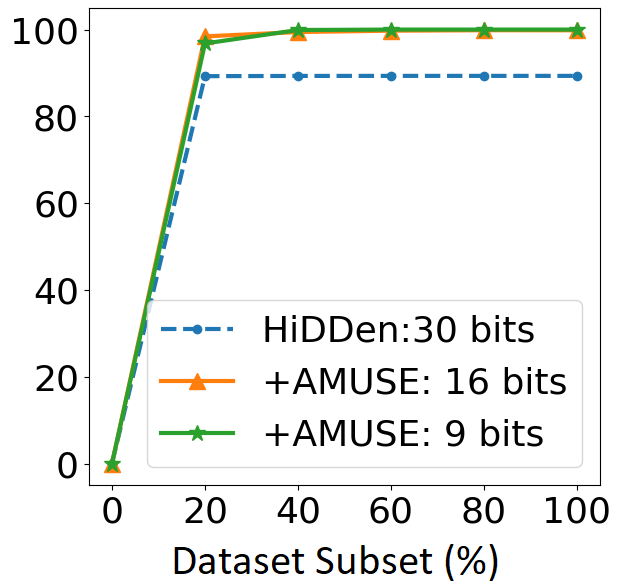}
\caption{The average bit accuracy after applying subset attack to SSL+AMUSE ($L=300$) and HiDDeN+AMUSE ($L=30$). 
}
\label{fig:subset_ssl_hidden_ba}
\end{figure}

\subsection{Is AMUSE robust to subset attack?}
As explained in the Problem Definition and Proposed Method sections, given the pre-defined threshold $\hat{\tau}$ (i.e., the subset ratio to guarantee correct message extraction from the watermarked dataset), AMUSE can accordingly adjust the sub-message length for both embedding and extraction.
So, when larger subsets are expected to be available during the extraction, the encoder distributes the message over larger number of samples resulting in shorter sub-messages.

In a subset attack, the adversary leaks a subset of the watermarked dataset to damage the extracted message. In this section, the robustness of the proposed AMUSE method is evaluated against subset attacks in terms of bit accuracy and word accuracy for different set of subset ratios. This study is performed using HiDDeN+AMUSE and SSL+AMUSE scenarios. 

For AMUSE, $\hat{\tau}=60\%$ and $\hat{\tau}=80\%$ are used to encode the original message into shorter sub-messages. 
When AMUSE is used, the original message is guaranteed to be reconstructed from any $\binom{N-1}{N-K}+1$ unique sub-messages  (assuming the sub-message extraction is 100\% correct). For $\hat{\tau}=60\%$ (with $N=5,K=3$), and $\hat{\tau}=80\%$ (with $N=5,K=4$), $\binom{N-1}{N-K}+1$ is equal to 7 and 5 unique sub-messages, respectively. The mentioned numbers are for the worst case scenario that guarantee error-free message reconstruction.
However, there are many cases where the original message can be reconstructed with less number of unique sub-messages. 
The subset attack analysis in this section is done on different subsets of randomly leaked samples including $s \in \{20,40,60,80,100\}\%$ with 100 trials for each $s$.

\begin{figure}[!tb]
\centering
\includegraphics[width=\columnwidth]{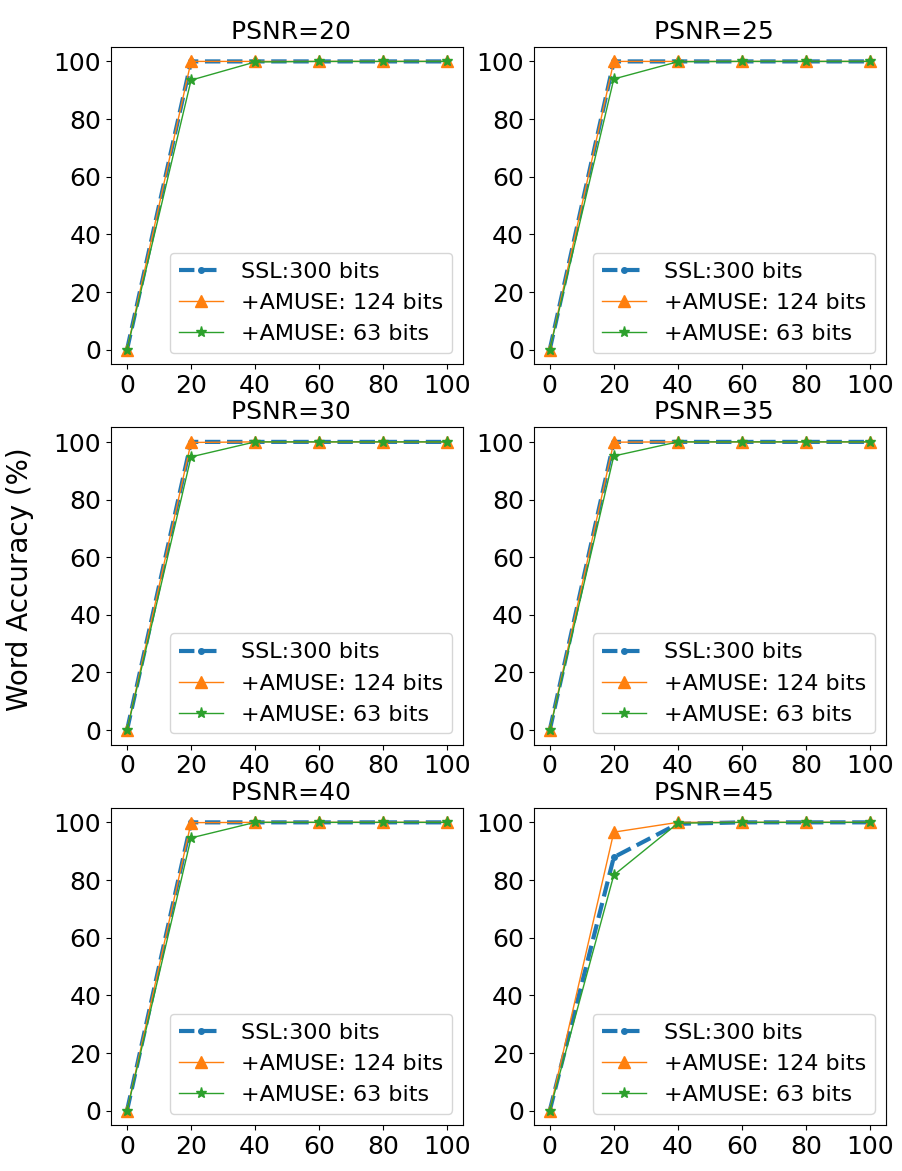}
\includegraphics[width=0.54\columnwidth]{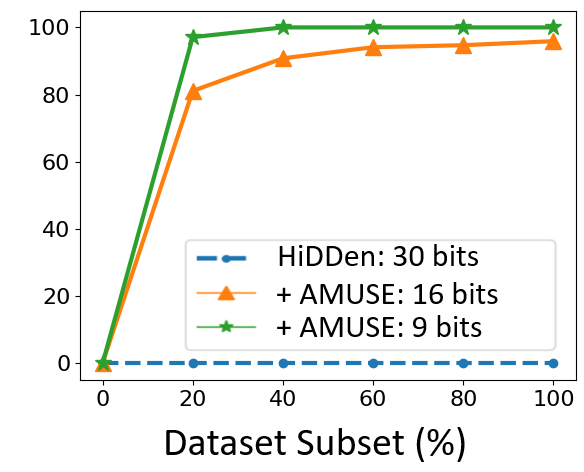}
\caption{The average word accuracy after applying subset attack to SSL+AMUSE (for different PSNR values with $L=300$) and HiDDeN+AMUSE (with $L=30$). }
\label{fig:subset_ssl_wa_300}
\end{figure}

For SSL+AMUSE, the average bit accuracy for no-attack case with $L$=300 and 6 PSNR values $\{20,25,30,35,40,45\}$ is shown in Fig.~\ref{fig:subset_ssl_hidden_ba} (top 6 plots). As shown in the figure, for all PSNR values, the proposed method is as accurate as the baseline for $s \geq 40\%$. For $s=20\%$, the bit accuracy dropped less than 4\% for $\hat{\tau}=80\%$ (i.e., 63-bit sub-message) compared to the baseline, while $\hat{\tau}=60\%$ (i.e., 124-bit sub-message) achieves the same results as the baseline. {The same pattern is observed for word accuracy that is provided in Fig.~\ref{fig:subset_ssl_wa_300}}. The bit and word accuracy plots for $L=\{100,200\}$ are given in {Appendix}.

For HiDDeN+AMUSE, the average bit extraction accuracy for no-attack case with $L=30$ is shown in Fig.~\ref{fig:subset_ssl_hidden_ba} (bottom plot). In this case, since applying AMUSE reduces the error in the reconstructed message, higher bit accuracy can be obtained compared to the baseline even at lower tested subset ratios.
{The corresponding word accuracy plots is provided in Fig.~\ref{fig:subset_ssl_wa_300}. Note that zero word accuracy means that there is at least one bit error in all the extracted messages.}

\section{Limitations}
In this section, we discuss some of the potential limitations and challenges of our method. One limitation of our work is that the extraction accuracy drops when severe attacks (e.g., very small crops, or very strong noise) which significantly damages the utility of a dataset are applied. However, low extraction accuracy for a damaged dataset is generally not a significant concern in practice. Hence, the high level of protection can be relaxed for such datasets due to their low utility. The other limitation is related to changing the required protection level $\hat{\tau}$ after watermarking. AMUSE watermarks a dataset for a given $\hat{\tau}$. If $\hat{\tau}$ changes, the dataset needs to be watermarked again with the updated parameter.

It should be noted that the message encoding/decoding steps of AMUSE are not significantly impacted by dataset size. However, watermarking all the samples in a large-scale dataset may add computational overhead, which depends on the efficiency of the watermarking algorithm not AMUSE. As a solution, one can use parallel processing for watermark embedding/extraction to efficiently handle watermarking of large-scale datasets.


\section{Conclusion}
\label{sec:conclude}
In this paper, we proposed AMUSE, an adaptive multi-segment message encoding-decoding method which maps the original watermark message into a set of shorter sub-messages and vice versa. Our encoder adaptively adjusts the length of the sub-messages to meet the target dataset protection requirements. Existing image watermarking methods are then utilized to embed and extract the sub-messages into and from the original images in the dataset. Then, our decoder is employed to reconstruct the original message from the extracted sub-messages. Our extensive experiments using different image watermarking methods showed that applying AMUSE improved the quality and the extraction accuracy (even with attacks such as subset attack) of the watermarked dataset. 

{While the focus of AMUSE is on image datasets, it is applicable to other modalities such as text and video. In the case of a text dataset, rather than embedding the entire watermark into every single text pieces (e.g., pages, paragraphs), the obtained sub-messages from AMUSE can be embedded into the text pieces. For a video dataset, rather than embedding the entire watermark message into all video files, the obtained sub-messages from AMUSE are embedded into the files in the dataset.  
In our future works, we will explore AMUSE for datasets of other modalities}.

\bibliography{aaai23}

\newpage 
\section{Appendix for AMUSE: Adaptive Multi-Segment Encoding for Dataset Watermarking}



\subsection{HiDDeN Training and Attacks}
The HiDDeN models were trained using 2.6K training samples from ImageNet training set. Three models with message lengths of 30, 16, and 9 bits were trained for 400 epochs. We set the batch size used for training all the models to 32. Some adversarial attacks were used during the training as an augmentation. The attacks that were considered during training include: Dropout (attack parameter $P$:  0.3 and 0.7), Cropout (attack parameter $P$: 0.7 and 0.3), Crop (attack parameter $P$: 0.7 and 0.3). Dropout and Cropout undo some of the watermarked pixels, resulting in an attacked image by combining pixels from the watermarked and the original images~\cite{hidden}. In both cases, $P\times100$\% of the pixels in the watermarked image were kept and the rest were replaced by the ones in the original image. In Dropout attack, the pixels in the watermarked images are chosen independently, while Cropout keeps a random square crop from the watermarked image. The crop attack produced a random square crop with the area equal to $P\times100$ of the watermarked image. The mentioned attacks are also applied to the test images (subset of ImageNet validation) and the average results are reported for with-attack experiments in the main material and the Appendix. 

\subsection{SSL attacks}
The list of 45 different attacks including transformations and noises which are applied to the watermarked images for SSL method is given in Table \ref{tbl:attacks}. The implementation details of each attack can be found in the Github repository of~\cite{ssl} \footnote{{\url{https://github.com/facebookresearch/ssl_watermarking/blob/main/evaluate.py}}}.

\subsection{Is AMUSE robust to subset attack?}

\begin{table}[!b]
    \renewcommand{\arraystretch}{0.80}
    \centering
\scalebox{0.8}{
\begin{tabular}{|cc||cc||cc|}
\hline
\textbf{attack} & \textbf{param} & \textbf{attack} & \textbf{param} & \textbf{attack}      & \textbf{param}       \\ \hline
blur         & 3         & contrast & 0.5       & resize   & 0.4       \\
blur         & 5         & contrast & 1.5       & resize   & 0.5       \\
blur         & 7         & contrast & 2.0       & resize   & 0.6       \\
blur         & 9         & hue      & -0.25     & resize   & 0.7       \\
blur         & 11        & hue      & -0.5      & resize   & 0.8       \\
blur         & 13        & hue      & 0.25      & resize   & 0.9       \\
brightness   & 0.5       & hue      & 0.5       & rotation & 5         \\
brightness   & 1.5       & jpeg     & 30        & rotation & 10        \\
brightness   & 2.0       & jpeg     & 40        & rotation & 15        \\
center\_crop & 0.4       & jpeg     & 50        & rotation & 20        \\
center\_crop & 0.5       & jpeg     & 60        & rotation & 25        \\
center\_crop & 0.6       & jpeg     & 70        & rotation & 30        \\
center\_crop & 0.7       & jpeg     & 80        & rotation & 35        \\
center\_crop & 0.8       & jpeg     & 90        & rotation & 40        \\
center\_crop & 0.9       & jpeg     & 100       & rotation & 45     \\ \hline
\end{tabular}
}
\caption{The list of attacks applied to the watermarked images (with the corresponding parameters) for SSL.}
\label{tbl:attacks}
\end{table}

\begin{table}[!tb]
\centering
\small
\begin{tabular}{c c c}
\toprule
\textbf{}   & \textbf{\begin{tabular}[c]{@{}c@{}}Dataset PSNR (dB)\end{tabular} } & \textbf{\begin{tabular}[c]{@{}c@{}} ML task \\ accuracy(\%)\end{tabular}} \\ \hline \hline
\begin{tabular}[c]{@{}c@{}}un-watermarked \ \end{tabular}        & Inf   & 91.97 \\ \hline 

\begin{tabular}[c]{@{}c@{}}HiDDeN\\ (30 bits/sample)\end{tabular}        & 30.40    & 90.98 \\ \hline 
\begin{tabular}[c]{@{}c@{}}HiDDeN+AMUSE \\ (16 bits/sample)\end{tabular} & 31.70    &  91.56 \\  \hline
\begin{tabular}[c]{@{}c@{}}HiDDeN+AMUSE \\ (9 bits/sample)\end{tabular}  & \textbf{32.13}    &  \textbf{91.69} \\ 
\bottomrule
\end{tabular}
\caption{Analysis of ML task results using Oxford Flower Dataset. PSNR: quality of watermarked dataset compared to the original unwatermarked dataset. Accuracy: the model performance after being trained on the given datasets.}
\label{tbl:flower}
\end{table}

\begin{figure*}[!tb]
\includegraphics[width=\columnwidth]{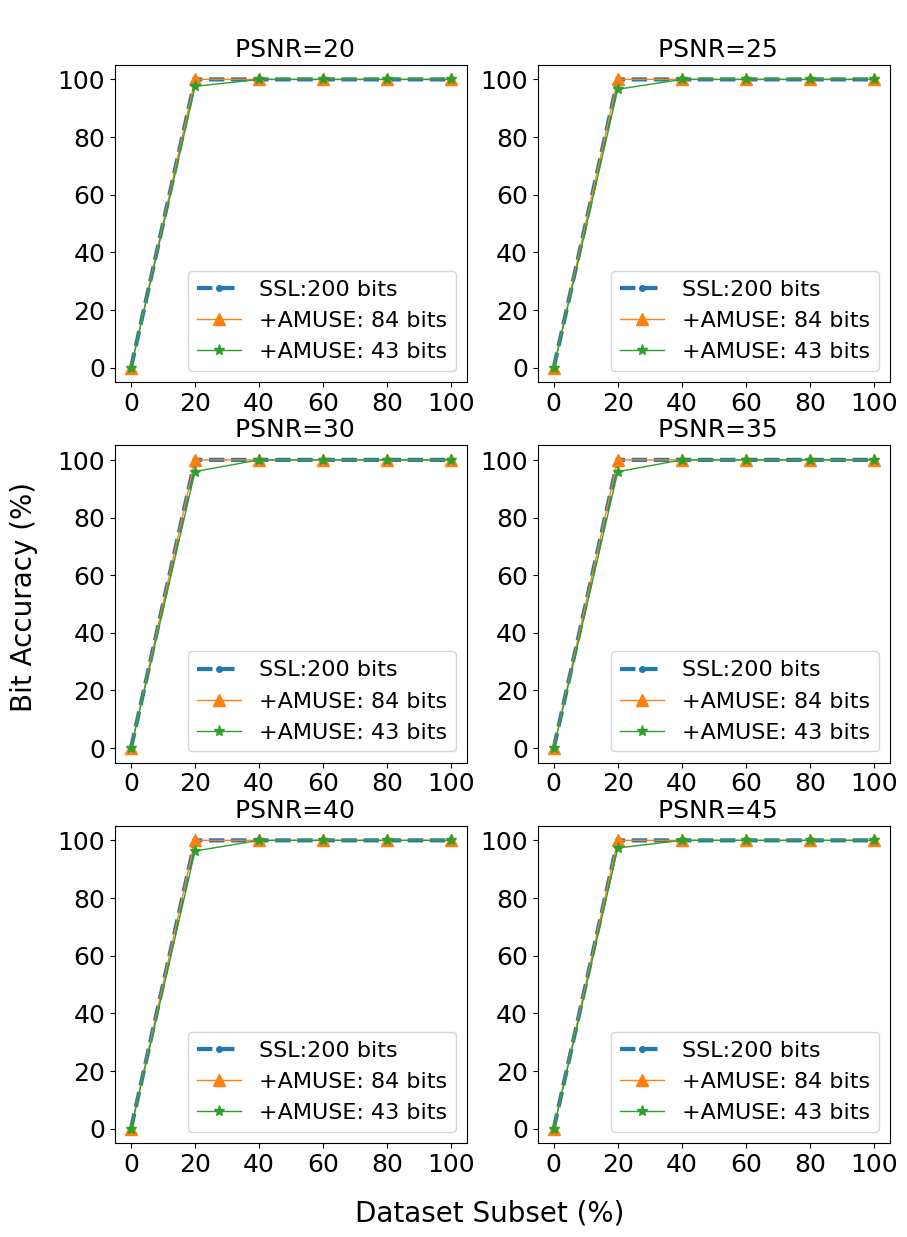}
\includegraphics[width=\columnwidth]{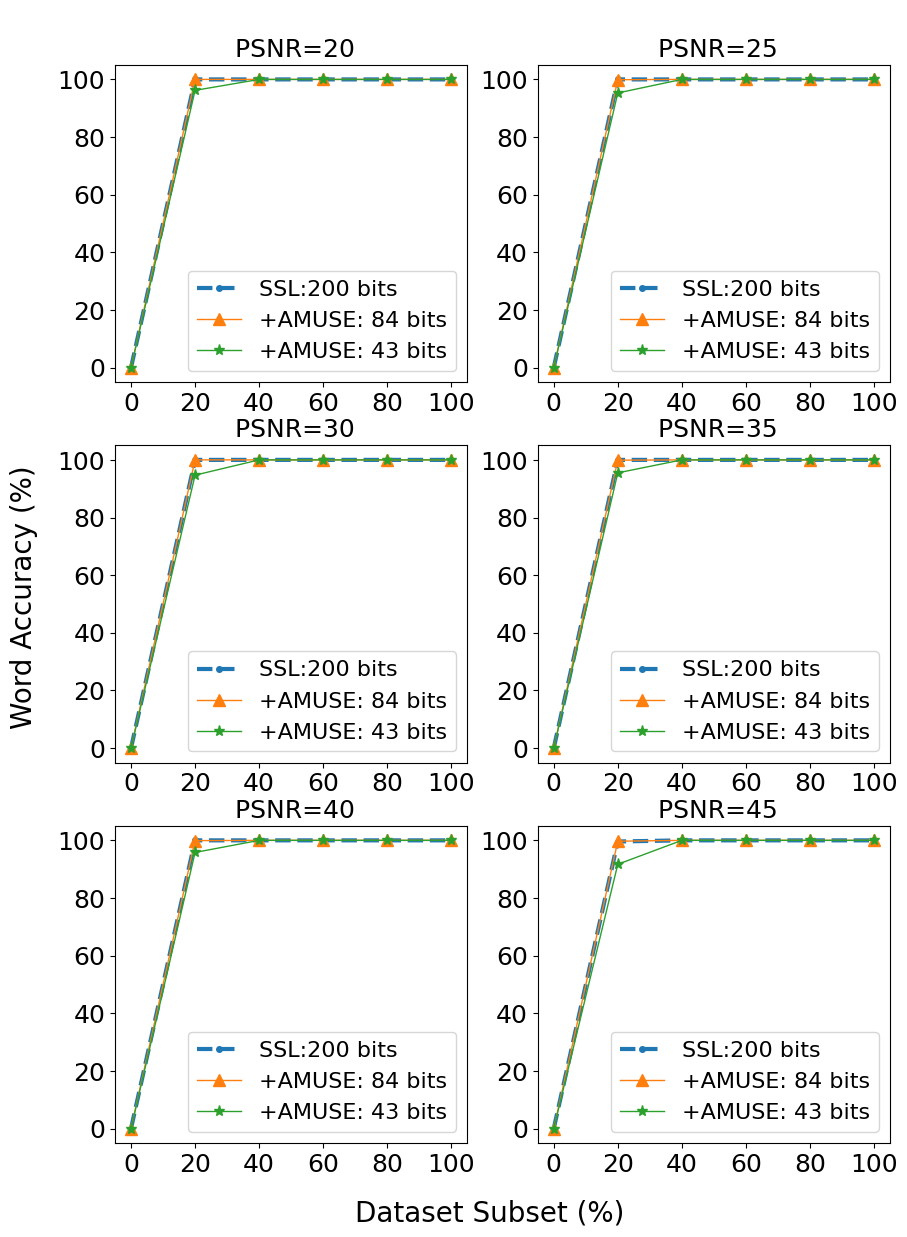}
\caption{The average bit accuracy (left two columns) and word accuracy (right two columns) after applying subset attack to SSL+AMUSE dataset watermarking for different PSNR values with $L=200$.}
\label{fig:subset_ssl_ba_200}
\end{figure*}

\begin{figure*}[!tb]
\includegraphics[width=\columnwidth]{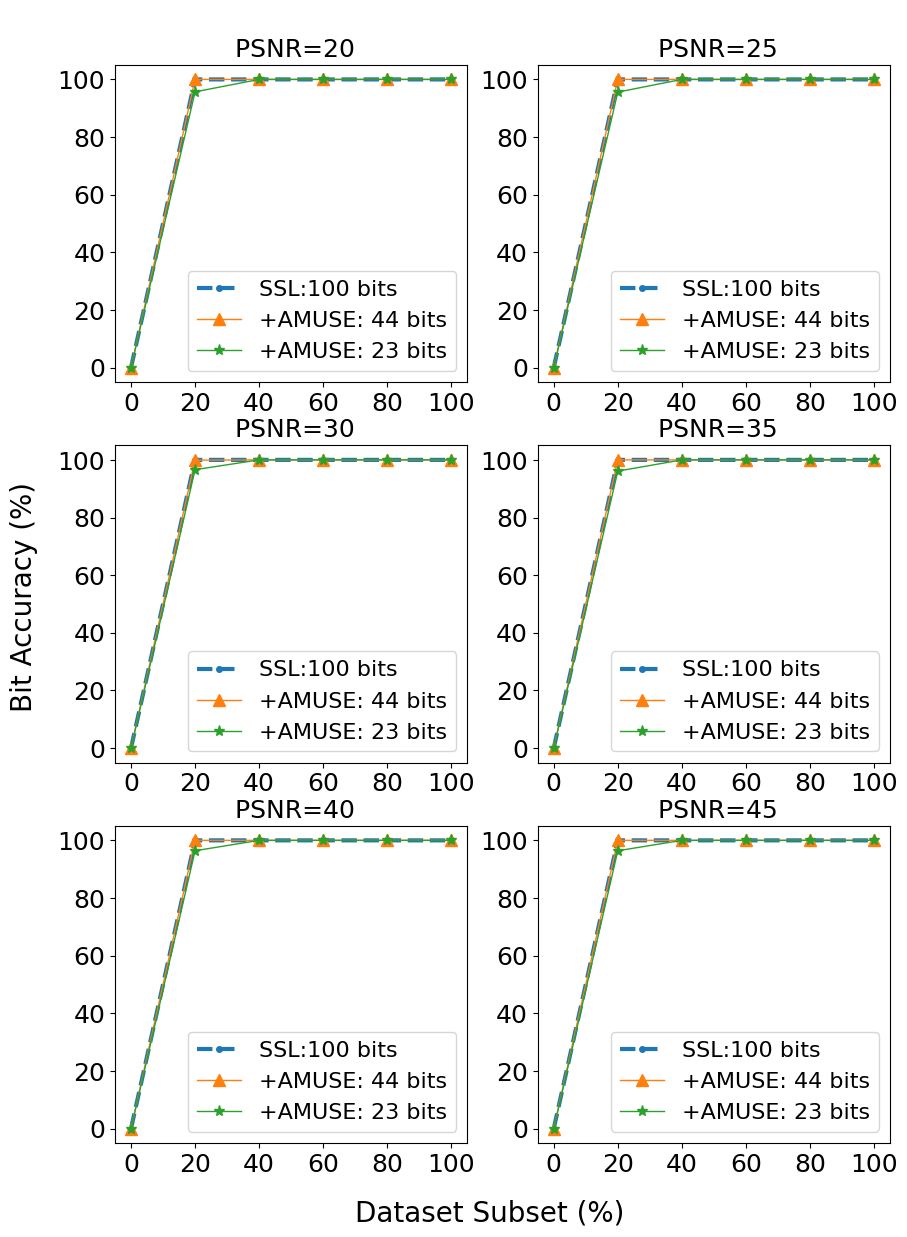}
\includegraphics[width=\columnwidth]{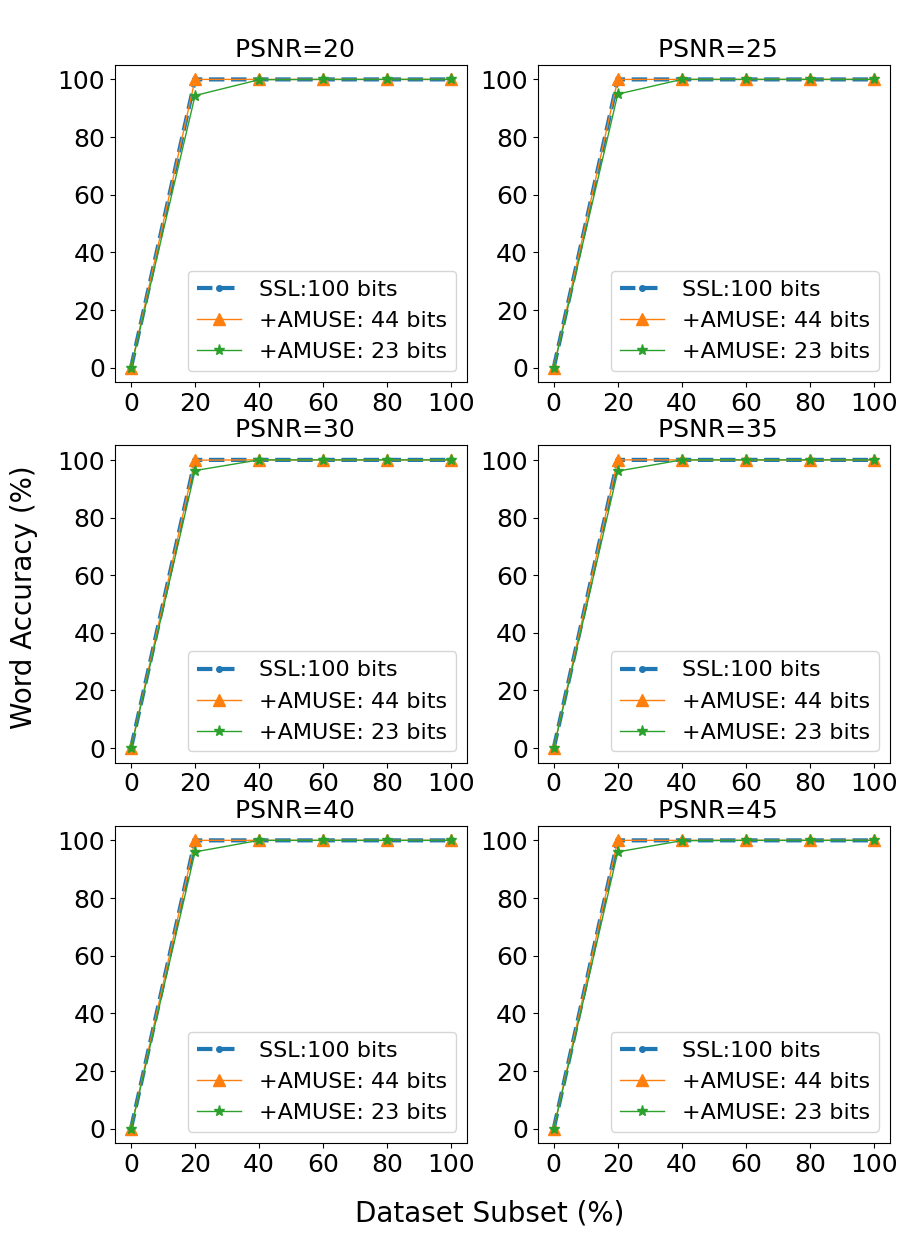}
\caption{The average bit accuracy (left two columns) and word accuracy (right two columns) after applying subset attack to the SSL+AMUSE dataset watermarking for different PSNR values with ${L=100}$.}
\label{fig:subset_ssl_ba_100}
\end{figure*}

\textbf{SSL+AMUSE}: 
In the paper, the  average  bit  accuracy  after  applying  subset  attack  to  SSL+AMUSE with $L=300$ was shown in Fig.~\ref{fig:subset_ssl_hidden_ba}. The corresponding word accuracies for the tested PSNR values were also shown here in Fig.~\ref{fig:subset_ssl_wa_300}. Here, the average bit and word accuracy after applying subset attack to SSL+AMUSE with $L=200$ and $L=100$ are presented in Fig.~\ref{fig:subset_ssl_ba_200} and Fig.~\ref{fig:subset_ssl_ba_100}, respectively. 
The results seen in the plots are aligned with the observed results in the main body of the paper.  

\begin{figure*}[!tb]
\centering
\includegraphics[width=\textwidth]{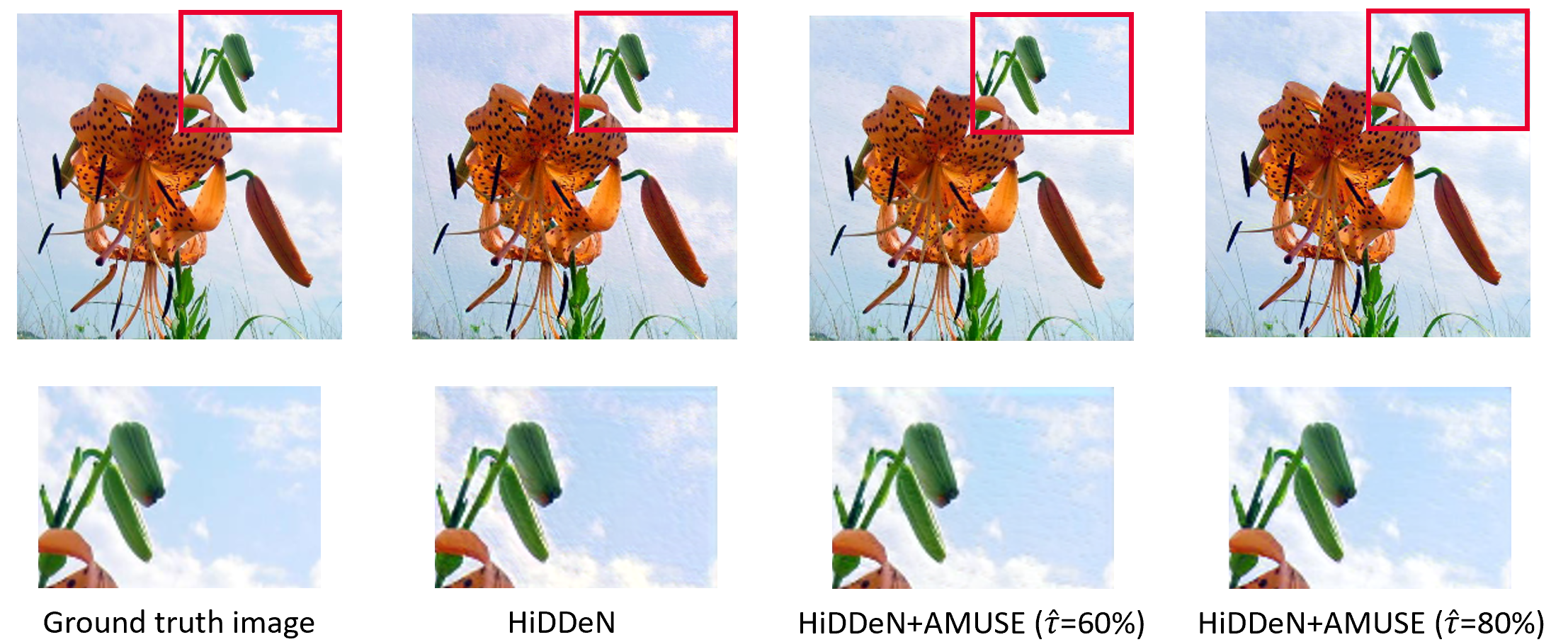}
\caption{Visual comparison between a sample image from the datasets watermarked using different methods with $L=30$. Top: full image, Bottom: zoomed image shown in the red rectangle on top.}
\label{fig:visual_example}
\end{figure*}

\subsection{Flower Dataset Results}
Oxford Flower Dataset~\cite{flower} was used to evaluate how watermarking a dataset affects the performance of a machine learning model that is trained on that dataset. To this end, HiDDeN+AMUSE dataset watermarking (presented in the paper) is used for watermarking the training set of Oxford Flower Dataset with $L=30$-bit messages. On the other hand, the validation and the test sets are left untouched. Following the work in~\cite{hidden}, we resize all the images to 128$\times$128 in this experiment. 

Four training sets are used to train a deep learning model. The first training set is the un-watermarked dataset which defines the anchor performance. The second dataset is the dataset which is watermarked by embedding the entire messages into all the samples of the dataset (i.e., the baseline in our experiments). The third and the fourth datasets are obtained by applying AMUSE with $\hat{\tau}=60\%$ and $\hat{\tau}=80\%$, respectively.  We obtained watermarked datasets with 10 different messages, and the average quality of the watermarked dataset compared to the original data is reported in Table~\ref{tbl:flower}. The comparison of the average PSNR values shows that applying AMUSE imporves the quality of the watermarked dataset in terms of PSNR. 

A visual example, comparing a sample in the mentioned training sets is shown in Fig.~\ref{fig:visual_example}. As shown in the figure, the distortion in the watermarked images becomes less noticeable when AMUSE is employed for dataset watermarking. As seen in the images, HiDDeN+AMUSE ($\hat{\tau}$=80) provides the highest visual quality with less artifacts (e.g., in the sky and clouds) compared to the others.

ResNet-34 models are trained using the obtained datasets to perform flower classification. The training is performed for 2500 iterations with batch size of 64. SGD optimizer with learning rate of 0.01, weight decay of 0.01, and CosineAnnealingLR~\cite{cosine} scheduler are utilized in this training procedure. 

After the models are trained, their performance is measured using the un-watermarked test set. Following~\cite{flower}, mean per class accuracy is used as the accuracy metric. The training is repeated 10 times using the watermarked training sets corresponding to 10 watermark messages. The average test accuracy over the trained models is reported in Table ~\ref{tbl:flower}. The test accuracy achieved using the un-watermarked training set is {91.97\%} which is the anchor test accuracy 
As it can be seen in the table, the test accuracy of the models trained on watermarked dataset drops when watermarked training set is used. However, HiDDeN+AMUSE achieves the highest test accuracy among the watermarked datasets. 




\end{document}